\documentclass[aps,preprint,prd,showpacs,nofootinbib]{revtex4}
\usepackage{comment}
\usepackage{graphicx}
\usepackage{dcolumn}
\usepackage{bm}
\usepackage{hyperref}
\usepackage{graphicx} 
\usepackage{subfigure}
\usepackage{bm}
\usepackage{latexsym}
\usepackage{ulem}
\usepackage{datetime}
\usepackage{scrtime}
\usepackage[dvipsnames, svgnames, x11names]{xcolor}

\usepackage{amsmath,amssymb,amsfonts,array}
\usepackage{bm,bbm}
\usepackage{cases}
\usepackage{dcolumn}
\usepackage{graphicx}
\usepackage{hyperref}
\usepackage{subfigure}
\usepackage{wasysym}
\usepackage{xcolor}

\definecolor{dullred}{rgb}{0.706,0.208,0.192}
\definecolor{darkred}{rgb}{0.545,0,0}
\definecolor{MaroonC}{rgb}{0,0.502,0.502}
\definecolor{dullblue}{rgb}{0,0.298,0.49}
\definecolor{blue3}{RGB}{31, 119, 180}
\definecolor{dullpurple}{rgb}{0.431,0.188,0.534}
\definecolor{darkgreen}{rgb}{0.075,0.302,0.047}
\definecolor{darkergreen}{rgb}{0,0.196,0.125}
\definecolor{darkergreen2}{rgb}{0,0.294,0.188}

\hypersetup{colorlinks=true,
	breaklinks=true,
	pdfstartview=Fit,
	linkcolor=blue,
	citecolor=blue,
	urlcolor=blue}

\begin{document}
\title{\boldmath Pulsar timing array observations as possible hints for nonsingular cosmology}	
	
\author{Mian Zhu$^{1,2}$}
\email[]{mzhuan@connect.ust.hk}
\author{Gen Ye$^{3}$}
\email[Corresponding author:~]{ye@lorentz.leidenuniv.nl}
\author{Yong Cai$^{1}$}
\email[Corresponding author:~]{caiyong@zzu.edu.cn}
\affiliation{$^1$ School of Physics and Microelectronics, Zhengzhou University, Zhengzhou, Henan 450001, China}
\affiliation{$^2$ Faculty of Physics, Astronomy and Applied Computer Science, Jagiellonian University, 30-348 Krakow, Poland}
\affiliation{$^3$ Institute Lorentz, Leiden University, PO Box 9506, Leiden 2300 RA, The Netherlands}

\begin{abstract}
Recent pulsar timing array (PTA) experiments have reported strong evidence of the stochastic gravitational wave background (SGWB). If interpreted as primordial gravitational waves (GWs), the signal favors a strongly blue-tilted spectrum. Consequently, the nonsingular cosmology, which is able to predict a strongly blue-tilted GW spectrum with $n_T \simeq 2$ on certain scales, offers a potential explanation for the observed SGWB signal. In this paper, we present a Genesis-inflation model capable of explaining the SGWB signal observed by the PTA collaborations while also overcoming the initial singularity problem associated with the inflationary cosmology. Furthermore, our model predicts distinctive features in the SGWB spectrum, which might be examined by forthcoming space-based gravitational wave experiments.
\end{abstract}

\maketitle
\flushbottom

\section{Introduction}

Inflation is the standard paradigm of the primordial Universe, providing a natural way to explain the formation of large scale structure and the observation of cosmic microwave background (CMB). Nonetheless, it is argued that inflation suffers from  the initial singularity problem \cite{Borde:1993xh,Borde:2001nh} (see, e.g., \cite{Lesnefsky:2022fen,Geshnizjani:2023edw} for recent development). Therefore, it is interesting to investigate nonsingular cosmologies, i.e. cosmological models that do not suffer the initial singularity problem, such as the bouncing cosmology \cite{Gasperini:1992em,Finelli:2001sr,Piao:2003zm,Piao:2004me,Piao:2005ag,Cai:2007qw,Cai:2011tc,Easson:2011zy,Cai:2012va,Liu:2013kea,Qiu:2013eoa,Koehn:2013upa,Wan:2015hya,Qiu:2015nha,Nojiri:2016ygo,Banerjee:2016hom,Chen:2017cjx,Mironov:2018oec,Akama:2018cqv,Ye:2019frg,Akama:2019qeh,Ye:2019sth,Mironov:2019mye,Nandi:2019xag,Nandi:2020szp,Battista:2020lqv,Zhu:2021whu,Banerjee:2022gpy,Ganz:2022zgs,Zhu:2023lhv,Singh:2023gxd,Burkmar:2023tll,Kaur:2023uaz,Tripathy:2023jid} and Genesis cosmology \cite{Creminelli:2010ba,Liu:2011ns,Wang:2012bq,Liu:2012ww,Creminelli:2012my,Hinterbichler:2012fr,Hinterbichler:2012yn,Easson:2013bda,Liu:2014tda,Pirtskhalava:2014esa,Nishi:2015pta,Cai:2016gjd,Nishi:2016ljg,Dobre:2017pnt,Mironov:2019qjt,Zhu:2021ggm,Cai:2022ori} (see also \cite{Piao:2003ty,Piao:2007sv}).

The standard slow-roll inflation predicts a nearly scale-invariant tensor spectrum with a tensor spectral index $n_T \simeq 0$.
On the contrary, many nonsingular scenarios predict strongly blue-tilted tensor spectra with $2\lesssim n_T \lesssim 3$ on certain scales \cite{Khoury:2001wf,Boyle:2003km,Qiu:2011cy,Nishi:2016wty}, provided the initial state of perturbation mode is set as the Bunch-Davis vacuum.
By introducing additional mechanisms during inflation, such as an intermediate null energy condition violation or a diminishment of propagating speed of Gravitational Waves (GWs), we can also obtain blue tensor spectra with $0<n_T \lesssim 2$ \cite{Piao:2004tq,Wang:2014kqa,Cai:2015yza,Cai:2016ldn,Wang:2016tbj,Cai:2020qpu,Cai:2022nqv}.
Therefore, the primordial gravitational wave (PGW) signal could be a promising tool for distinguishing canonical slow-roll inflation and its competitors or modifications, including certain nonsingular cosmologies.

Recently, pulsar timing array (PTA) collaborations, including NANOGrav \cite{NANOGrav:2023hvm,NANOGrav:2023gor}, EPTA \cite{Antoniadis:2023ott}, PPTA \cite{Reardon:2023gzh}, and CPTA \cite{Xu:2023wog}, have reported strong evidence for an isotropic stochastic GW background with a strain amplitude of order $\mathcal{O}(10^{-15})$ at the reference frequency $f = 1\, \textnormal{yr}^{-1}$. The PTA result strongly supports a blue tensor spectrum with $n_T = 1.8 \pm 0.3$ (see e.g., \cite{NANOGrav:2023hvm,Antoniadis:2023zhi,Vagnozzi:2023lwo}). This value of $n_T$ still falls within the range predicted by various primordial cosmological scenarios (see \cite{Piao:2004jg} for the spectra index of various expanding and contracting phases). It is then natural to ask if it is possible to interpret the PTA signals to be originated from PGWs in nonsingular cosmology.\footnote{The PTA results can of course have different origins. Shortly after the release of the PTA data, an abundant paper appears and tries to explain the PTA signals, see, e.g., \cite{Huang:2023chx,Madge:2023cak,Ellis:2023dgf,Ellis:2023tsl,Franciolini:2023pbf,Zhang:2023lzt,Cannizzaro:2023mgc,Zhang:2023nrs,Balaji:2023ehk,Du:2023qvj,Zhu:2023faa,Shen:2023pan,DiBari:2023upq,Li:2023bxy,Li:2023tdx,Konoplya:2023fmh,Niu:2023bsr,Ghosh:2023aum,Wang:2023ost,Wu:2023hsa,Liu:2023ymk,Liu:2023pau,Jiang:2023qbm,Cai:2023dls,Addazi:2023jvg,Bian:2023dnv,Han:2023olf,Guo:2023hyp,Ye:2023xyr,Jiang:2023gfe,Choudhury:2023kam,Cheung:2023ihl,Oikonomou:2023qfz}, see also e.g. \cite{Sakharov:2021dim,Ashoorioon:2022raz,Lazarides:2023ksx,Odintsov:2021kup,Oikonomou:2022irx,Oikonomou:2023bah}.}

In the minimal setting, nonsingular cosmologies like Ekpyrotic bounce and Genesis cosmology predict blue tensor spectra on all scales, which is not realistic: the power spectrum of tensor perturbation (i.e., $P_T$) is either too small to be observed on the PTA scale or grows too large on smaller scales to invalidate the perturbation theory. Additionally, for a power law like $P_T\sim k^{n_T}$ in inflationary scenario, there is a very low upper limit on the reheating temperature (i.e., $T_\mathrm{rh}\lesssim 10$ GeV \cite{Vagnozzi:2023lwo}) due to the constraint on the $e$-folding number of inflation implied by $P_T<1$. A realistic candidate for an early universe model might be a combination of nonsingular cosmology and inflation, which is able to naturally terminate the blue nature of $P_T$ on certain scales as well as resolve the cosmological singularity problem.


As a start, we begin with a Genesis-inflation model \cite{Cai:2017tku}, due to its relative simplicity. We construct a specific Genesis-inflation model such that the scalar spectrum is scale-invariant in both the quasi-Minkowskian epoch and the inflation epoch, consistent with CMB observations. The tensor spectrum is blue-tilted across a broad frequency range from the observational window of CMB to that of PTA, and scale-invariant on smaller scales, with oscillatory features on a short range of scales corresponding to the transition epoch. We find that the PGW generated in this model can successfully explain the PTA data, while its tail in the higher frequency band might be accessible to future GW detectors such as LISA \cite{LISA:2017pwj}, Taiji \cite{Hu:2017mde} and TianQin \cite{TianQin:2015yph}.


This paper is organized as follows. In Sec. \ref{sec:model}, we introduce our toy model of Genesis-inflation. We analyze the dynamics of perturbation in Sec. \ref{sec:pt}, then numerically evaluate the tensor perturbation and confront it with PTA observations in Sec. \ref{sec:num}. We conclude in Sec. \ref{sec:conclusion}.
We take the sign of the metric as $(-,+,+,+)$ throughout. The canonical kinetic term of the scalar field $\phi$ is defined as $X \equiv -\frac{1}{2}\nabla_{\mu}\phi \nabla^{\mu} \phi$, which is simply $X = \dot{\phi}^2/2$ at the background level. The d'Alembert operator is $\Box \equiv \nabla_{\mu}\nabla^{\mu}$. We have also set $M_p\equiv(8\pi G)^{-1/2}=1$ for simplicity.

\section{A simple Genesis model}
\label{sec:model}
\subsection{Action}
\label{sec:action}

The ``no-go'' theorem, as proven in \cite{Libanov:2016kfc, Kobayashi:2016xpl}, indicates that spatially flat nonsingular cosmologies constructed within Horndeski theories are plagued by ghost or gradient instabilities. However, it has been demonstrated within the framework of effective field theory (EFT) that these instabilities can be eliminated through the use of ``beyond Horndeski'' operators \cite{Cai:2016thi, Creminelli:2016zwa, Cai:2017tku, Cai:2017dyi, Kolevatov:2017voe} (see also \cite{Mironov:2019qjt, Ilyas:2020zcb, Zhu:2021ggm, Cai:2022ori} for developments in Genesis cosmology). Notably, it is demonstrated in \cite{Cai:2017tku} with the least set of EFT operators that a fully stable nonsingular Genesis-inflation model can be implemented.

In this paper, we start with the action
\begin{equation}
	\label{eq:action}	
	S = \int d^4x \sqrt{-g} \left( \frac{R}{2} + \mathcal{L}_{H2} + \mathcal{L}_{H3} + \mathcal{L}_\mathrm{EFT} \right) ~,
\end{equation}
where $R/2$ is the standard Einstein-Hilbert action, $\mathcal{L}_{H2}$ and $\mathcal{L}_{H3}$ are the Horndeski Lagrangian. Specifically, we take
\begin{equation}
	\mathcal{L}_{H2} = g_1(\phi)  X + g_2(\phi) X^2 - V(\phi) ~,\quad~ \mathcal{L}_{H3} = -\gamma X^2 \Box \phi ~.
\end{equation}
The EFT Lagrangian $\mathcal{L}_\mathrm{EFT}$ in (\ref{eq:action}) represents the ``beyond Horndeski'' operators, which are able to stabilize the scalar perturbations. As we primarily focus on primordial GWs in this paper, we will not delve into the details of $\mathcal{L}_\mathrm{EFT}$, see e.g. \cite{Cai:2017tku}.


In a spatially flat FLRW universe
\begin{equation}
    ds^2 = -dt^2 + a(t)^2 d\vec{x}^2 ~,
\end{equation}
the Friedmann equations are
\begin{equation}
\label{eq:Friedmann1}	
	3H^2 = \rho_{\phi} = \frac{1}{2} g_1 \dot{\phi}^2 + \frac{3}{4} g_2 \dot{\phi}^4 + 3H\gamma \dot{\phi}^5 + V(\phi) ~,
\end{equation}
\begin{equation}
    \dot{H} = - \frac{1}{2} \left( \rho_{\phi} + p_{\phi} \right) = -\frac{1}{2}g_1 \dot{\phi}^2 - \frac{1}{2}g_2 \dot{\phi}^4 - \frac{1}{2} \gamma \dot{\phi}^4 (3\dot{\phi} H - \ddot{\phi}) ~.
\end{equation}
where $H$ is the Hubble parameter, $\rho_{\phi}$ and $p_{\phi}$ are the density and pressure of the matter field $\phi$. An overdot represents the differentiation with respect to the physical time $t$.

As we will show in Sec. \ref{sec:Gbg}, a Genesis solution with a scale-invariant scalar spectrum can be realised by the following choice of auxiliary functions
\begin{equation}
\label{eq:g}
    g_1(\phi) = -\frac{3\kappa^2e^{4\phi}}{2} \frac{1 - \tanh \omega_1 (\phi - \phi_1)}{2} ~,\quad~ g_2(\phi) = e^{2\phi} \frac{1 + e^{2\phi}}{1 + e^{4\phi}} ~,~
\end{equation}
\begin{equation}
\label{eq:V}
    V(\phi) = \frac{1}{2}m^2 \left( 1 - \frac{\phi^2}{\phi_0^2} \right)^2 \left( \frac{1 + \tanh \omega_2 (\phi - \phi_2)}{2}  \right) ~.
\end{equation}

\subsection{The Genesis solution}
\label{sec:Gbg}
Assuming the universe adopts a Genesis solution in the past infinity with $\phi \to -\infty$, we expect the auxiliary functions to have the following asymptotic behavior:
\begin{equation}
\label{eq:glimit-inf}	
	\lim_{\phi \to -\infty} g_1(\phi) = -\frac{3}{2}\kappa^2 e^{4\phi} ~,~ \lim_{\phi \to -\infty} g_2(\phi) = e^{2\phi} ~,~  \lim_{\phi \to -\infty} V(\phi) = 0 ~.
\end{equation}
Moreover, to have a late-time slow-roll inflationary epoch, we need a flat potential:
\begin{equation}
\label{eq:glimit+inf}	
	\lim_{\phi \to +\infty} g_1(\phi) = 0 ~,~ \lim_{\phi \to +\infty} g_2(\phi) = 1 ~,~  \lim_{\phi \to +\infty} V(\phi) = \frac{1}{2}m^2 \left( 1 - \frac{\phi^2}{\phi_0^2} \right)^2 ~.
\end{equation}
Therefore, we choose the explicit functions \eqref{eq:g} and \eqref{eq:V} to smoothly connect the asymptotic forms \eqref{eq:glimit-inf} and \eqref{eq:glimit+inf}.

The Friedmann equations in the far past, with the asymptotic behavior \eqref{eq:glimit-inf}, gives
\begin{equation}
\label{eq:GH2}
    3H^2 = -\frac{3}{4} \kappa^2 e^{4\phi} \dot{\phi}^2 + \frac{3}{4} e^{2\phi} \dot{\phi}^4 + 3H\gamma \dot{\phi}^5 ~,
\end{equation}
which permits a quasi-Minkowskian solution
\begin{equation}
\label{eq:Genesisbg}
    \dot{\phi}^2 e^{-2\phi} = \kappa^2 ~\to~ \phi = \ln \left[ \frac{1}{\kappa(-t)} \right] ~,~ \dot{\phi} = \frac{1}{-t} ~,~ e^{\phi} = \frac{1}{\kappa (-t)} ~.
\end{equation}
The second Friedmann equation reads
\begin{equation}
\label{eq:GdotH}
    \dot{H} \simeq \frac{1 + 2\kappa^2 \gamma}{4\kappa^2} \frac{1}{(-t)^6} ~\to~ H \simeq \frac{1 + 2\kappa^2 \gamma}{20\kappa^2} \frac{1}{(-t)^5} > 0 ~.
\end{equation}
We will use \eqref{eq:Genesisbg} and \eqref{eq:GdotH} as the initial condition for the numerical evaluation.

\subsection{Illustration of background dynamics}
\begin{figure}[htp]
    \centering
    \includegraphics[width=0.3\linewidth]{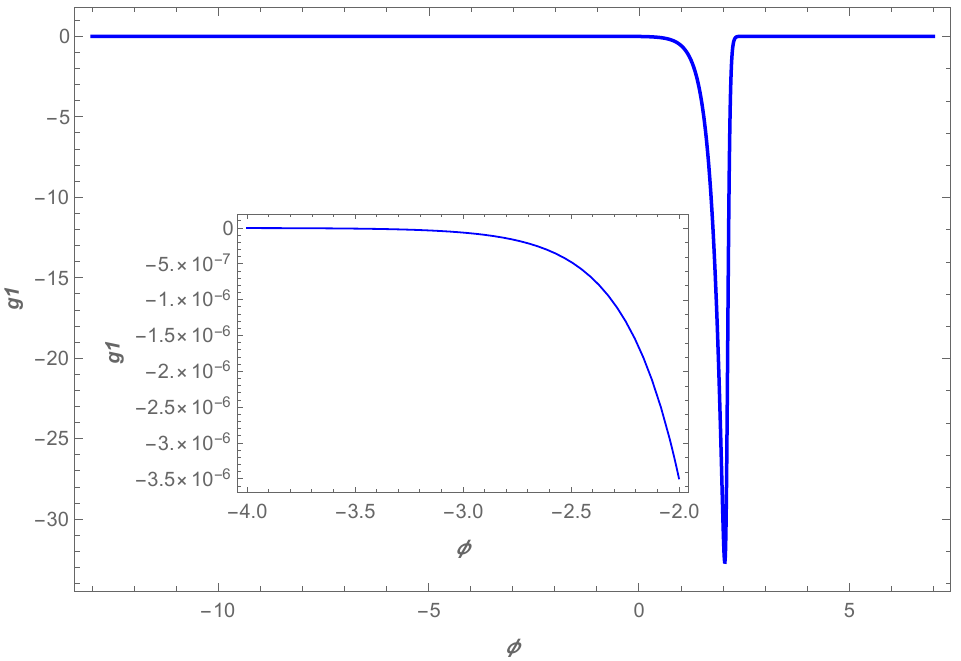}
    \includegraphics[width=0.3\linewidth]{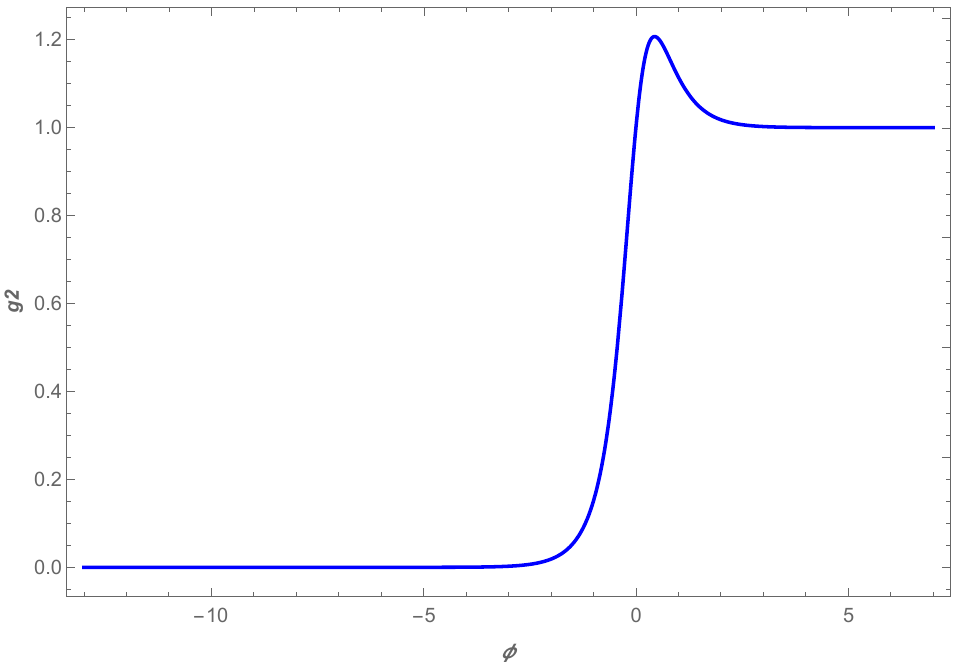}
    \includegraphics[width=0.31\linewidth]{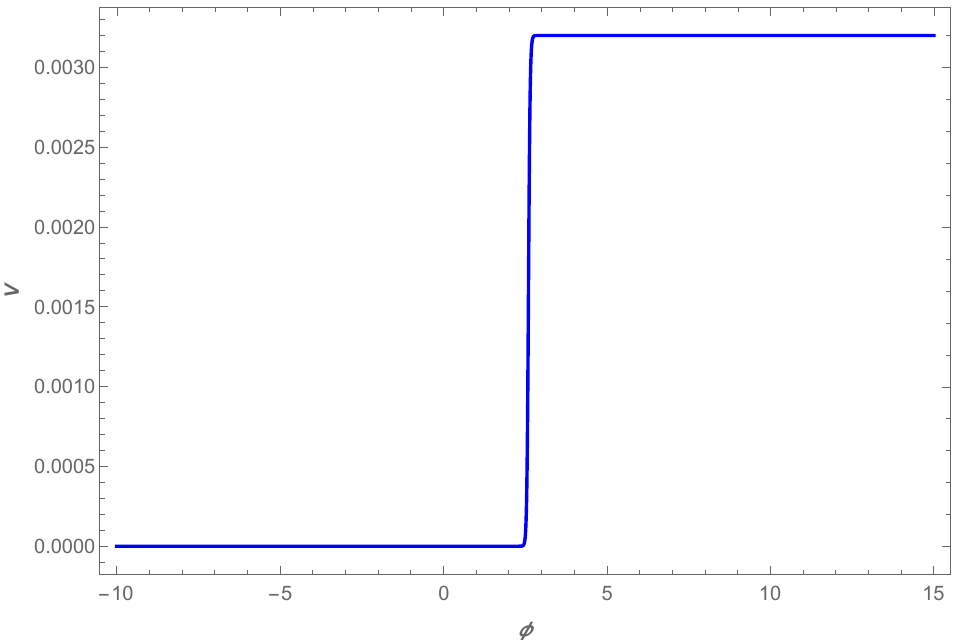}
    \caption{The auxiliary functions.}
    \label{fig:gV}
\end{figure}

We numerically evaluate the background dynamics in this section. We adopt the parameter setting
\begin{equation}
    \label{eq:para1}
    \kappa = \frac{1}{12} ~,~ \gamma = 72 ~,~ m = 0.08 ~;
\end{equation}
\begin{equation}
\label{eq:para2}
    \omega_1 = 18 ~,~ \omega_2 = 20 ~,~ \phi_0 = 2000 ~,~ \phi_1 = 2.1 ~,~ \phi_2 = 2.6 ~.
\end{equation}
The parameter set \eqref{eq:para2} ensures the asymptotic behavior \eqref{eq:glimit-inf} and \eqref{eq:glimit+inf}, and has little influence on the relevant physics. We plot the auxiliary functions $g_1$, $g_2$ and $V$ in Fig. \ref{fig:gV}.

On the other hand, the parameters \eqref{eq:para1} determine the physics of Genesis cosmology. From \eqref{eq:GdotH} we see that $\kappa$ and $\gamma$ determine the Hubble parameter in the quasi-Minkowskian epoch. Moreover, in Sec. \ref{sec:spt} we will show that the sound speed of scalar perturbation $c_s^2$ in the quasi-Minkowskian epoch is also determined by $\kappa$ and $\gamma$, and we choose \eqref{eq:para1} such that $c_s^2 = 1$.\footnote{In nonsingular cosmology with a consistency relation between the tensor-to-scalar $r$ and $c_s^2$ like matter bounce, interpreting the PTA data would result in an extremely small $r$, and thus a highly suppressed $c_s^2$, which in turns give an unsuppressed non-Gaussianity, inconsistent with observations \cite{Vagnozzi:2023lwo}. In our case however, $r$ depends on other parameters as well as $c_s^2$, so it's possible to acquire the required $r$ with $c_s^2 = 1$.} The parameter $m$ determines the height of ``cliff'' in the inflationary potential $V(\phi)$, and thus the Hubble parameter in the inflation stage.

\begin{figure}[htp]
    \centering
    \includegraphics[width=0.4\linewidth]{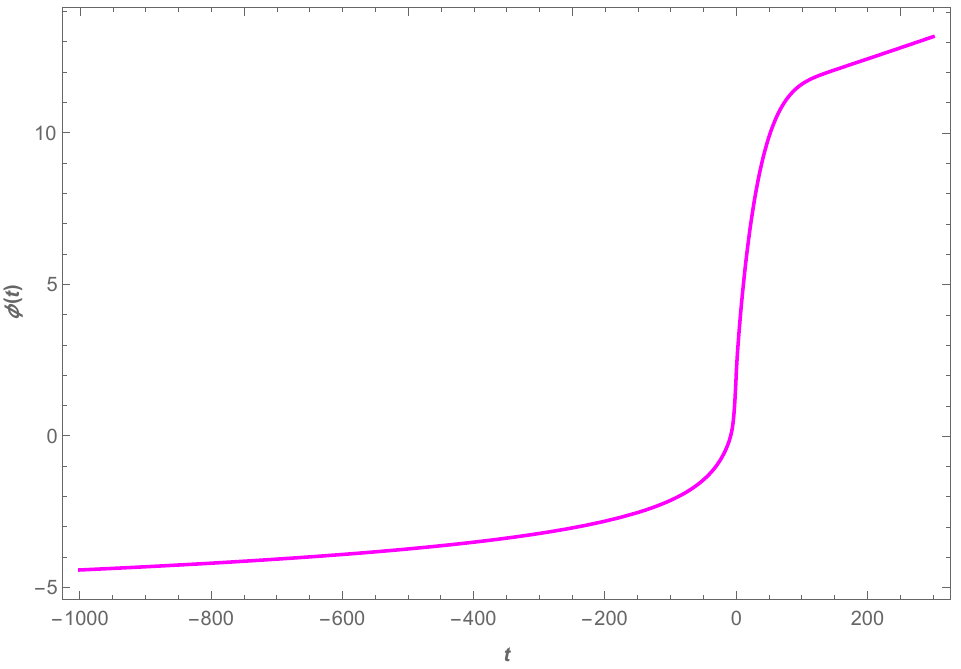}
    \includegraphics[width=0.4\linewidth]{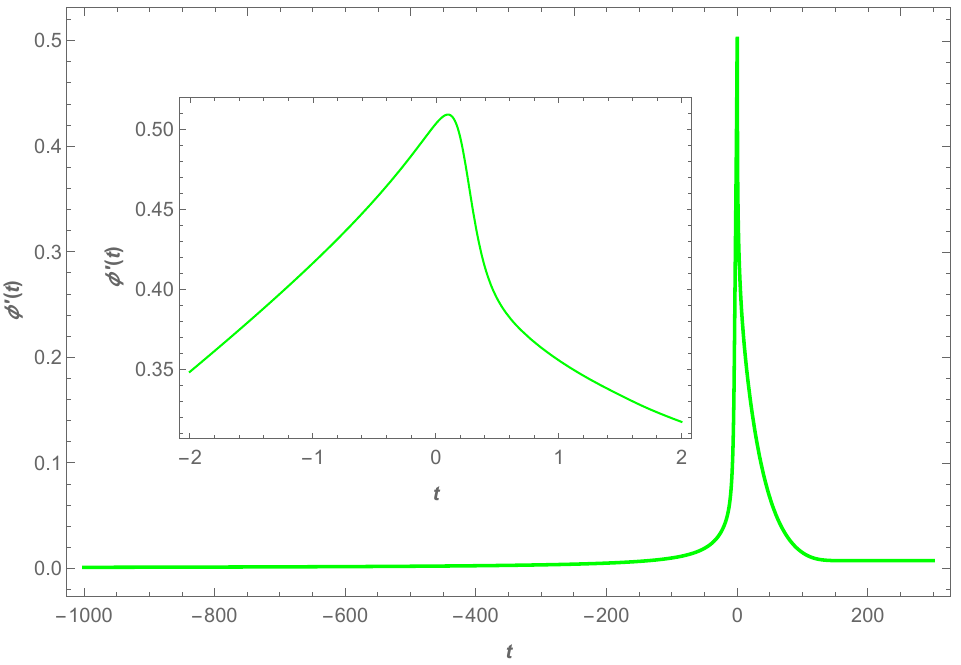} \\
    \includegraphics[width=0.4\linewidth]{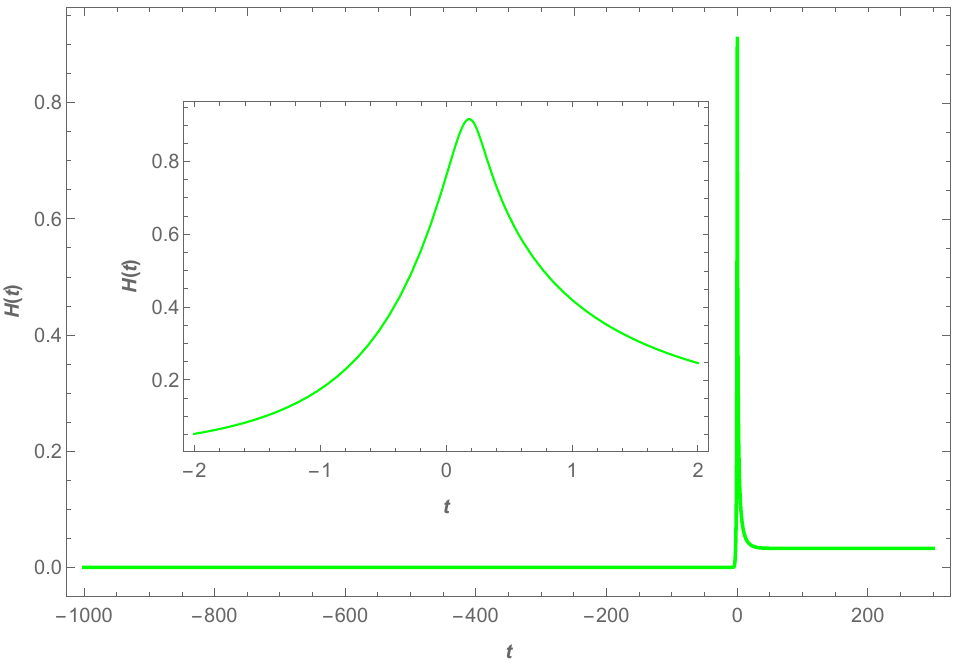}
    \includegraphics[width=0.4\linewidth]{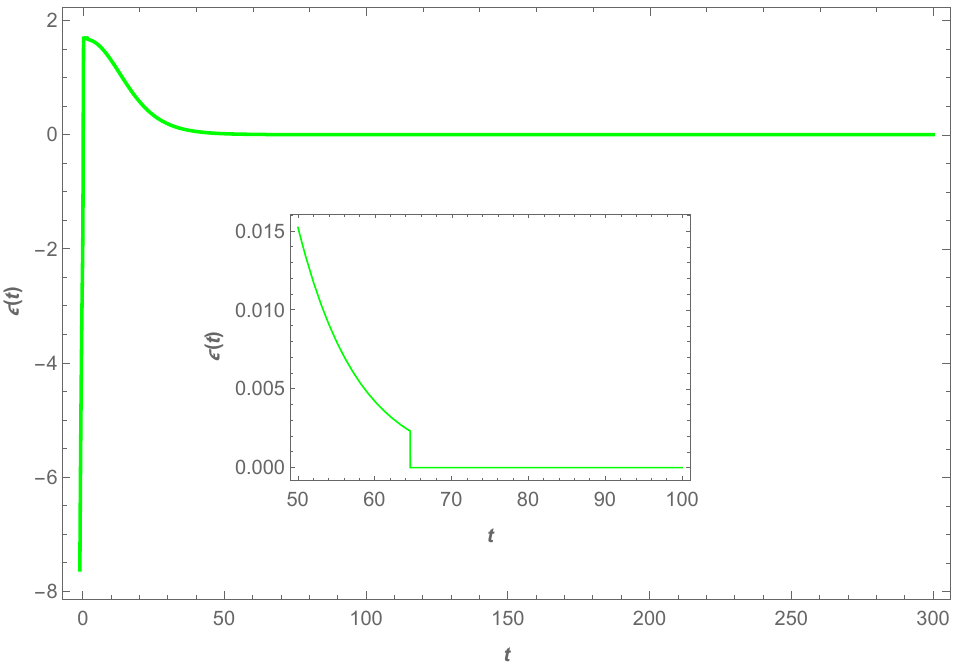}
    \caption{The background dynamics of our Genesis-inflation model with parameter setting \eqref{eq:para1}, \eqref{eq:para2}. The upper panel shows the dynamics of the scalar field, while the lower panel shows the dynamics of the background evolution.}
    \label{fig:bg}
\end{figure}

We illustrate the background dynamics of our model in Fig. \ref{fig:bg}. The quasi-Minkowskian epoch is characterized by the $H \to 0$ behavior when $t<0$. After a short transition epoch around $t=0$, the universe enters an inflation epoch with a nearly constant $H$.

\section{Perturbation}
\label{sec:pt}

\subsection{Dynamic equation}
\label{sec:Ms}
The scalar and tensor perturbations at the quadratic level can be expressed as
\begin{equation}
\label{eq:S2s}	
	S_{2,s} = \int d\tau d^3x \frac{z_s^2}{2} \left[ \zeta^{\prime 2} - c_s^2 (\partial_i \zeta)^2 \right] ~,
\end{equation}
\begin{equation}
	S_{2,T} = \int d\tau d^3x \frac{a^2}{8} \left[ \gamma_{ij}^{\prime 2} - c_T^2 \gamma_{ij}^2 \right] ~,
\end{equation}
where $\zeta$ is the curvature perturbation and $\gamma_{ij}$ represents the tensor perturbation. Here for simplicity, we suppress the polarization tensor for the tensor sector. We have
\begin{equation}
\label{eq:zs2def}
	\frac{z_s^2}{2a^2} = 3 + 2\frac{\dot{\phi}^2(g_1 + 3g_2\dot{\phi}^2) + 18H\gamma \dot{\phi}^5 - 6H^2}{(2H-\gamma \dot{\phi}^5 )^2} ~,
\end{equation}
and
\begin{equation}
	\left( - \frac{z_s^2}{2a^2} \right) c_s^2 = 1 - \frac{2}{a} \frac{d}{dt} \left(\frac{a}{2H - \gamma \dot{\phi}^5} \right) = \frac{\gamma^2 \dot{\phi}^{10} - 2\gamma H \dot{\phi}^5 - 10 \gamma \dot{\phi}^4 \ddot{\phi} + 4\dot{H}}{(2H - \gamma \dot{\phi}^5)^2} ~.
\end{equation}
in the quasi-Minkowskian epoch and the inflation epoch \cite{Ilyas:2020qja}.

Notably, the behavior of $c_s^2$ is dependent on the EFT action in the transition epoch. However, in our case we are interested in the scale-invariance behavior: the scale invariance in the quasi-Minkowskian epoch can be used to interpret the CMB result, while the scale invariance in the inflation epoch guarantees the scalar perturbation does not grow on small scales and invalidate the perturbation theory. Therefore, let's simply assume that the EFT action is delicately designed, such that $c_s^2 > 0$ in the transition period, and no gradient instability occurs.

The dynamical equation of the scalar and tensor perturbation is
\begin{equation}
	\label{eq:MSscalar}
	v_k^{\prime \prime} + \left(k^2c_s^2 - \frac{z_s^{\prime \prime}}{z_s}\right) v_k = 0 ~,
\end{equation}
\begin{equation}
	\label{eq:MStensor}
	\nu_k^{\prime \prime} + \left(c_T^2k^2 - \frac{a^{\prime \prime}}{a} \right) \nu_k = 0 ~,
\end{equation}
where $v_k \equiv z_s \zeta$ and $\nu_k \equiv 1/2 a\gamma_k$ are the corresponding mode functions and a prime denotes differentiation with respect to the conformal time $d\tau = dt/a$.

\subsection{Effective horizon}
Since the expression $z_s^2$ in the quasi-Minkowskian epoch, \eqref{eq:zs2def}, is non-trivial, let's define the effective ``Hubble horizon'' for scalar and tensor perturbation:
\begin{equation}
\label{eq:HEff}
    H_s^2 \equiv z_s^{\prime \prime}/ z_s ~,~ H_T^2 \equiv a^{\prime \prime} / a ~.
\end{equation}
From the dynamical equations \eqref{eq:MSscalar} and \eqref{eq:MStensor}, we see the effective Hubble parameters determine the evolution of perturbations.

Let's first come to the tensor mode. The effective Hubble parameter behaves as
\begin{equation}
\label{eq:HT2}
    \lim_{t \to -\infty} H_T^2 = a^2(2H^2 + \dot{H}) \propto (-t)^{-6} \simeq 0 ~,~ \lim_{t \to \infty} H_T^2 = a^2H^2 (2 - \epsilon) \simeq \frac{2}{\tau^2} ~.
\end{equation}
Therefore, the tensor perturbation will remain in the vacuum state in the quasi-Minkowskian epoch, and thus the tensor spectra index is simply $n_T = 2$. In the subsequent inflationary epoch, modes crossing the effective horizon would be redshifted and get a nearly scale-invariant tensor spectrum with $n_T = 0$, as predicted by the standard slow-roll inflation.

Denoting the minimal value of $H_T^2$ in the inflationary epoch to be $H_{T,\min}$, we can estimate
\begin{equation}
    n_T = 2 ~,~ k < k_T ~;~ n_T = 0 ~,~ k > k_T ~;~ k_T \equiv H_{T,\min} ~.
\end{equation}
The behavior of $H_S^2$ is more complicated. However, as we shall elaborate in Sec. \ref{sec:spt}, for modes crossing the effective Hubble horizon during the quasi-Minkowskian epoch and the inflationary epoch, their corresponding scalar spectra index is $n_s \simeq 1$. Thus, we expect the scalar spectrum to be almost scale-invariant, except for a possible feature at scales corresponding to the transition epoch.

\subsection{Tensor spectra}
\label{sec:tpt}
In the quasi-Minkowskian epoch, $H_T^2$ is approximately zero, so the dynamical equation is simply
\begin{equation}
    \mu_k^{\prime \prime} + k^2 \mu_k = 0 ~\to ~ |\mu_k| = \frac{1}{\sqrt{2k}} ~,
\end{equation}
after imposing the vacuum initial condition. The corresponding tensor spectrum is
\begin{equation}
\label{eq:PTG}
    P_T \equiv \frac{4k^3}{\pi^2} \frac{|\mu_k|^2}{a^2} \simeq  \frac{2}{\pi^2} \left(\frac{k}{a_0}\right)^2 ~,
\end{equation}
where $a_0$ represents the scale factor during the Genesis epoch, $k/a_0$ is the physical wavenumber. Apparently, for the modes that exit the horizon during the Genesis epoch, the tensor spectral index is $n_T=2$.

During the inflationary epoch, the dynamic equation takes the form
\begin{equation}
\mu_k^{\prime \prime} + \left( k^2 - \frac{2}{(\tau - \tau_e)^2} \right) \mu_k = 0 ,
\end{equation}
where $\tau_e$ is an integration constant. Denoting the start time of inflation as $\tau_I$, we find
\begin{equation}
k_T \equiv H_{T,\min} = \frac{\sqrt{2}}{|\tau_I - \tau_e|} ~.
\end{equation}

For modes with $k < k_T$, they are already superhorizon at the start of inflation, and thus their amplitude remains constant. The corresponding spectrum is described by \eqref{eq:PTG}, where $a=a(\tau_M)$, with $\tau_M$ denoting the end of the quasi-Minkowskian state.
On the other hand, for modes with $k>k_I$, the general solution with vacuum initial conditions is given by
\begin{equation}
\mu_k = \frac{\sqrt{{\pi} |\tau - \tau_e|}}{2} H_{3/2}^{(1)} (k |\tau - \tau_e|) = \frac{-i - k|\tau - \tau_e|}{\sqrt{2} k^{3/2} |\tau - \tau_e|} e^{ik|\tau - \tau_e|} ~,
\end{equation}
where $H_{\nu}^{(1)}$ represents the Hankel function of the first kind.

Utilizing the inflationary background given by
\begin{equation}
a(\tau) = \frac{1}{H_I |\tau - \tau_e|} ~,
\end{equation}
where $H_I \simeq \textnormal{Const}$ represents the Hubble parameter during the inflationary epoch, the corresponding tensor spectrum becomes
\begin{equation}
P_T \equiv \frac{4k^3}{\pi^2} \frac{|\mu_k|^2}{a^2} = \frac{2H_I^2}{\pi^2} ~,
\end{equation}
at the super-horizon scale $k |\tau - \tau_e| \ll 1$.

In conclusion, we estimate the features of the tensor spectrum as follows
\begin{equation}
P_T = \frac{2}{\pi^2} \frac{k^2}{a(\tau_M)^2} , k < k_T ; P_T = \frac{2H_I^2}{\pi^2} , k > k_T ~.
\end{equation}

\subsection{Scalar perturbation}
\label{sec:spt}
In the quasi-Minkowskian epoch, the universe is almost static, and $d\tau = dt/a \propto dt$. By properly redefining the conformal time, we can interchangeably use $a_0 \tau$ and $t$. Moreover, the EFT operator is negligible, along with the asymptotic behavior \eqref{eq:glimit-inf} and \eqref{eq:Genesisbg}, the parameters reduce to
\begin{equation}
	\label{eq:zscst<0}
	z_s^2 = a_0^2 \frac{600 \kappa^2}{(8\kappa^2 \gamma - 1)^2}(-t)^4 = \frac{600 \kappa^2 a_0^6}{(8\kappa^2 \gamma - 1)^2}(-\tau)^4 ~,~ c_s^2 = \frac{8\kappa^2\gamma-1}{3} \equiv c_{s0}^2 ~,
\end{equation}
where we keep only the leading term. Equation \eqref{eq:MSscalar} then becomes
\begin{equation}
	v_k^{\prime \prime} + \left(k^2c_{s0}^2 - \frac{2}{\tau^2}\right)v_k = 0 ~,
\end{equation}
whose solution, equipped with the vacuum initial condition, is
\begin{equation}
	\label{eq:vktau}
	v_k(\tau) = \frac{\sqrt{{\pi} (-\tau)}}{2} H_{3/2}^{(1)} (-k c_{s,0} \tau) ~,
\end{equation}
and for $-k \tau \ll 1$, we have
\begin{equation}
    |v_k (\tau)| = \frac{1}{\sqrt{2}k^{3/2} c_{s,0}^{3/2} (-\tau)} ~,
\end{equation}
Hence, the scalar power spectrum can be evaluated by
\begin{equation}
\label{eq:Pzeta}	
	P_{\zeta}(k) = \frac{k^3}{2\pi^2} \frac{|v_k|^2}{z_s^2} = \frac{(8\kappa^2 \gamma - 1)^2}{2400\pi^2 c_{s,0}^3 \kappa^2 a_0^6 (-\tau)^6} = \frac{3 c_{s,0}}{800\pi^2 \kappa^2 (-t)^6} ~.
\end{equation}

Notice that, the quasi-Minkowskian epoch ends at a specific time $\tau_M < 0$. It is pointed out in \cite{Liu:2011ns,Nishi:2016ljg} that, for $\mathcal{L}_{H3} \propto X^{\alpha} \Box \phi$ with $\alpha > 1/2$, the curvature perturbation grows on super-horizon scales, which is consistent with our result \eqref{eq:Pzeta}. In view of that, we shall use the end time of the quasi-Minkowskian epoch, $t_M$, to evaluate the amplitude of scalar spectrum.\footnote{Following the convention of \cite{Liu:2011ns}, we have $Q\simeq\Lambda_*^4(-t)^4$, where $\Lambda_*^4\simeq100\kappa^2/(3c_{s,0}^2)$. The end time of the quasi-Minkowskian epoch can be approximately evaluated with $Q\simeq1$. As a result, we find $t_M\simeq -(c_{s,0}/\kappa)^{1/2}$. With Eq. (\ref{eq:Pzeta}), we find $P_{\zeta}\sim \kappa/c_{s,0}^2$, which is nearly scale-invariant, for the perturbation modes exiting horizon during the Genesis epoch. For simplicity, we will not delicately design the magnitude of the scalar power spectrum in this paper.}
The modes with $-k \tau_M < 1$ will cross the horizon and have the power spectrum described by \eqref{eq:Pzeta}, while modes with $-k \tau_M > 1$ remain sub-horizon in the whole quasi-Minkowskian epoch.

In the inflationary epoch, modes with $-k \tau_M \gg 1$ will cross the horizon and acquire a scale-invariant spectrum. The only tricky mode is $-k \tau_M \simeq 1$, whose evolution is hard to trace analytically. Fortunately, the corresponding modes are in a small range of scales, which shall generate features in the power spectrum for a limited $k$. Thus the scalar power spectrum is almost scale-invariant, as expected.

One final issue remains to be addressed. To ensure numerical robustness, we adopted a very flat potential \eqref{eq:para1}, leading to a slow-roll parameter $\epsilon \ll 1$ during the inflationary epoch (see Fig. \ref{fig:bg}). Unfortunately, in canonical inflation, the consistency relation dictates that $r \propto \epsilon$, resulting in an extremely small $r$ on small scales. Consequently, the scalar spectrum exhibits a magnitude much larger than unity, leading to the breakdown of perturbation theory. While our current manuscript focuses on the possibility of the PTA signal being a hint of nonsingular cosmology and mainly emphasizes the tensor spectrum, we should address this issue in future works involving a concrete realization of nonsingular scenarios. This could be achieved by either constructing an inflation epoch with moderate $\epsilon$ or employing non-canonical kinetic terms to break the consistency relation.

\section{Numerical evaluation}
\label{sec:num}

In this section, we will perform numerical evaluations of the primordial tensor perturbations and compare them with the most up-to-date PTA data. We will utilize the specific parameter settings given by \eqref{eq:para1} and \eqref{eq:para2}. Additionally, to compare our results with observations, we need to transform the primordial tensor spectrum into the spectral energy density parameter observed today. For simplicity, we assume that the inflationary epoch is followed by instantaneous reheating, as well as the standard radiation, matter, and dark energy eras. Consequently, the spectral energy density parameter $\Omega_{GW}(k)$ is related to the primordial tensor spectrum $P_T$ by
\begin{equation}
\Omega_{\textnormal{GW}}(k) \simeq 10^{-6} P_T(k) ~.
\end{equation}

Besides, we need to specify the scale factor in \eqref{eq:PTG}. The scale factor $a_0$ in the Genesis epoch is related to today's scale factor $a_{\text{today}} = a_0 e^{N}$, where $N$ is the e-folding number from the end of Genesis epoch to today. Insert back the $M_p$'s which has been set to unity, the tensor spectrum \eqref{eq:PTG} is accordingly
\begin{equation}
    P_T = \frac{2}{\pi^2} \frac{k^2}{a_{\text{today}}^2M_p^2} e^{2N} \simeq 5.8\times 10^{-103} e^{2N}\left(\frac{f}{\rm{nHz}}\right)^2 ~,
\end{equation}
To explain the PTA results, we need $P_T(\rm{nHz})\sim 10^{-3}$, which implies $N \simeq 114$. \footnote{This might lead to the trans-Planckian problem at high frequency \cite{Brandenberger:2000wr,Martin:2000xs}.}
\begin{figure}[htp]
    \centering
    \includegraphics[width=0.7\linewidth]{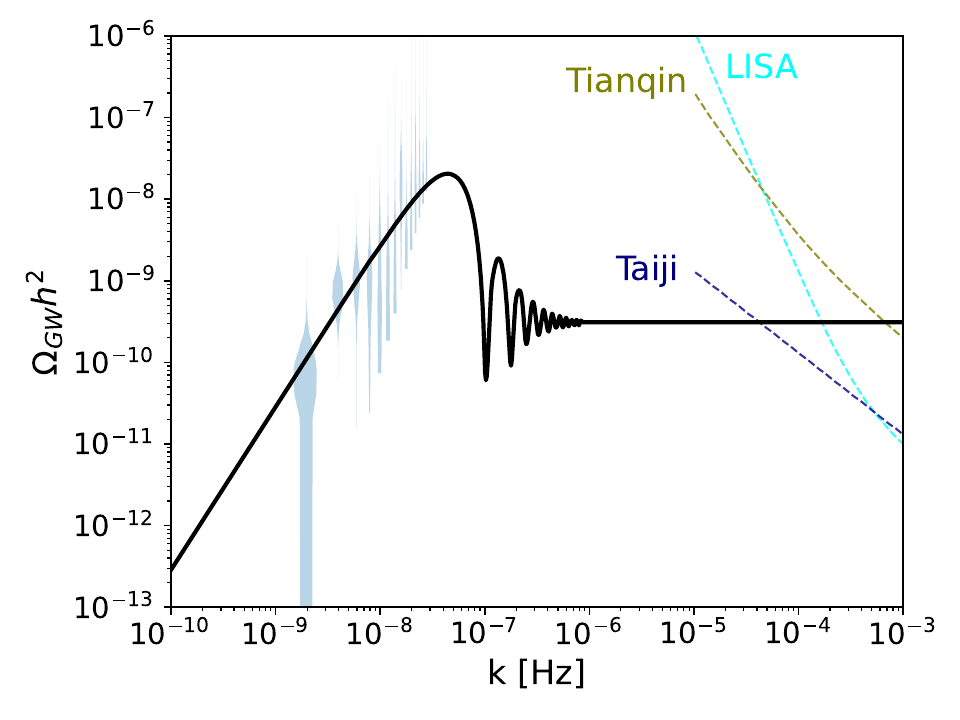}
    \caption{The spectral energy density parameter predicted by our model. The violin represents the NANOGrav result, and the dashed curves on the right represent the sensitivity of ongoing space-based GW detectors, including LISA \cite{LISA:2017pwj}, Taiji \cite{Hu:2017mde} and TianQin \cite{TianQin:2015yph}.}
    \label{fig:OGW}
\end{figure}

We present the spectral energy density parameter in Fig. \ref{fig:OGW}. Our model not only exhibits a blue spectrum on the nHz scales, explaining the recent NanoGrav data, but also generates significant PGW signals on smaller scales, which might be probed by upcoming space-based GW detectors such as LISA, Taiji and TianQin. This distinctive feature could set our scenario apart from others. For instance, in the scalar-induced gravitational waves interpretation of PTA data \cite{Domenech:2019quo, Domenech:2021ztg}, the peaked gravitational wave signals are typically redshifted on scales smaller than the PTA scale, leading to undetectable signals on smaller scales. Consequently, our scenario could be distinguished from others through future experiments.

\section{Conclusion and Outlook}
\label{sec:conclusion}

We examine whether the most up-to-date  PTA results may offer insights into nonsingular cosmology.
To achieve this, we investigate a toy Genesis-inflation model described by the action \eqref{eq:action}, which is able to yield a nearly scale-invariant scalar power spectrum. The tensor spectrum exhibits a blue tilt with $n_T = 2$ over a wide frequency range, spanning from the observational window of the CMB to that of PTA. This characteristic allows for a direct comparison with PTA observations. Additionally, the amplitude of GWs could be substantial enough to be detectable by forthcoming space-based GW detectors, making our scenario potentially testable in the near future.



In this paper, we explore the aforementioned possibility through a toy Genesis-inflation model. There are many interesting questions to address in the forthcoming studies.

We need to account for scalar perturbations in the model. In Section \ref{sec:spt}, we argue that the inflation epoch must be carefully treated to ensure a sizable PGW signal on the PTA scale while keeping the scalar perturbation small. This necessitates a relatively large tensor-to-scalar ratio $r$ or a breakdown of the consistency relation in canonical inflation. Thus, we will pay meticulous attention to the inflation (and the transition) epoch to ensure the safety of the scalar perturbation.



Additionally, we interpret the PTA observation as a result of amplified PGWs. As pointed out by \cite{Inomata:2022yte}, if the primordial scalar spectrum is amplified to certain scales, the back reaction at a non-linear level might be too strong to break the perturbation theory. Although the study focuses on scalar perturbation, a similar issue can potentially arise in our scenario. More specifically, to explain the PTA result through amplified PGWs, the tensor spectrum must have a minimal amplitude on PTA scales and smaller scales. It is essential to address, whether this minimal amplitude will always result in model-independent large back-reaction, or the back-reaction issue is dependent on the model construction.


Furthermore, we find that in our toy model, there is a potential trans-Planckian issue at high frequency band. It's also known that Genesis cosmology might suffer from the strong coupling problem \cite{Ageeva:2020buc, Ageeva:2021yik, Akama:2022usl}. Further examination of healthier realization in nonsingular scenarios, such as the Ekpyrosis-bounce-inflation scenario, is necessary.

\section*{ACKNOWLEDGMENTS}
We are grateful to Yi-Fu Cai, Chunshan Lin, Yun-Song Piao, Taotao Qiu, Yi Wang for stimulating discussions. M.Z. is supported by grant No. UMO 2021/42/E/ST9/00260 from the National Science Centre, Poland. G.~Y. is supported by NWO and the Dutch
Ministry of Education, Culture and Science (OCW) (grant VI.Vidi.192.069). Y. C. is supported in part by the National Natural Science Foundation of China (Grant No. 11905224), the China Postdoctoral Science Foundation (Grant No. 2021M692942) and Zhengzhou University (Grant No. 32340282).

\bibliography{reference}
\bibliographystyle{apsrev4-1}

\end{document}